\documentclass[convention,express-paper]{aesconf}

\graphicspath{{./}{figures/}}

\usepackage[utf8]{inputenc}

\usepackage{microtype}

\usepackage[numbers,square]{natbib}

\usepackage{booktabs}
\usepackage{color}
\usepackage{url}
\usepackage{soul}
\usepackage{amsmath}
\usepackage{amssymb}
\title{Decoding Vocal Articulations from Acoustic Latent Representations}

\author[1,2]{Mateo Cámara}
\author[1]{Fernando Marcos}
\author[1,2]{José Luis Blanco}

\affil[1]{Grupo de Aplicaciones del Procesado de Señal, Universidad Politécnica de Madrid, Spain}
\affil[2]{Information Processing and Telecommunications Center, Universidad Politécnica de Madrid, Spain}

\correspondence{Mateo Cámara}{mateo.camara@upm.com}

\lastnames{Cámara, Marcos, and Blanco}

\shorttitle{Vocal Articulations from Acoustic Representations}

\savebox{\AEStop}{%
	\begin{minipage}{\textwidth}%
		\rule{\textwidth}{1.5pt}\\%
		\\%
		\begin{minipage}[c][\iftoggle{convention}{3.2cm}{3.7cm}][t]{0\textwidth}%
			\includegraphics[width=20mm]{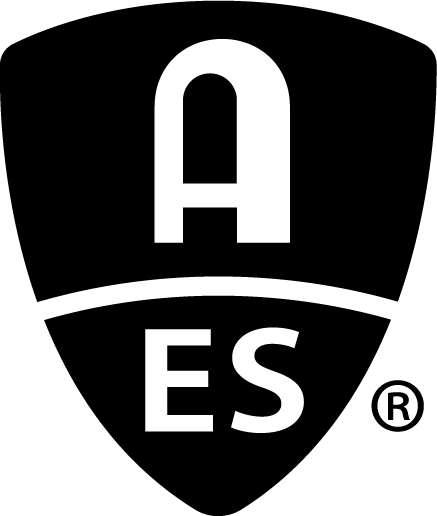}%
		\end{minipage}%
		\begin{minipage}{\textwidth}%
			\sffamily%
			\begin{center}%
				\LARGE Audio Engineering Society\\%
				\iftoggle{express_paper}{%
				\hspace{3mm}\fontsize{36}{38pt}\selectfont Convention Express\\Paper \AESExpressPaperNumber\\%
				}{%
				\iftoggle{convention}{%
				\fontsize{36}{38pt}\selectfont Convention Paper\\%
				}{%
				\fontsize{36}{38pt}\selectfont Conference Paper\\%
				}}%
				\vspace{0.2cm}%
				\large Presented at the AES \ifx\AESConferenceNumber\empty\else\AESConferenceNumber \fi\iftoggle{convention}{Convention\\}{\AESConferencePrefix Conference on\\}%
				\iftoggle{convention}{}{\AESConferenceTopic\\}%
				\AESConferenceDate\ifx\AESConferenceLocation\empty\else, \AESConferenceLocation\fi%
			\end{center}%
		\end{minipage}\\%
		\vspace{0.2cm}\\%
		\begin{minipage}{\textwidth}%
			\rmfamily\itshape\small	\AESLegalTextPrefix\ \AESLegalText%
		\end{minipage}\\%
		\\%
		\rule{\textwidth}{1.5pt}%
	\end{minipage}%
}

\begin{document}

\twocolumn[
\maketitle 

\begin{onecolabstract}
We present a novel neural encoder system for acoustic-to-articulatory inversion. We leverage the Pink Trombone voice synthesizer that reveals articulatory parameters (e.g tongue position and vocal cord configuration). Our system is designed to identify the articulatory features responsible for producing specific acoustic characteristics contained in a neural latent representation. To generate the necessary latent embeddings, we employed two main methodologies. The first was a self-supervised variational autoencoder trained from scratch to reconstruct the input signal at the decoder stage. We conditioned its bottleneck layer with a subnetwork called the "projector," which decodes the voice synthesizer's parameters.

The second methodology utilized two pretrained models: EnCodec and Wav2Vec. They eliminate the need to train the encoding process from scratch, allowing us to focus on training the projector network. This approach aimed to explore the potential of these existing models in the context of acoustic-to-articulatory inversion. By reusing the pretrained models, we significantly simplified the data processing pipeline, increasing efficiency and reducing computational overhead.

The primary goal of our project was to demonstrate that these neural architectures can effectively encapsulate both acoustic and articulatory features. This prediction-based approach is much faster than traditional methods focused on acoustic feature-based parameter optimization. We validated our models by predicting six different parameters and evaluating them with objective and ViSQOL subjective-equivalent metric using both synthesizer- and human-generated sounds. The results show that the predicted parameters can generate human-like vowel sounds when input into the synthesizer. We provide the dataset, code, and detailed findings to support future research in this field.
\end{onecolabstract}
]

\section{Introduction}

Exploring the complexities inherent in human speech production is a cornerstone linguistic research challenge. An avenue to address this challenge involves delving into the acoustic signals emanating from the human vocal tract. These signals offer insights into various levels of speech and their correlation with established acoustic cues. One can dissect this process using physically informed synthesis models. The acoustic-to-articulatory inversion shall allow us to infer the configuration of the articulators, while a physically informed synthesizer should recreate the audio records.


Artificial Intelligence emerges as a promising tool for tackling this task. Amalgamating diverse representations, we aim to deepen our understanding of the process. A plausible way is to combine representations aimed at speech analysis, synthesis, or recognition to estimate a speech articulation model that mimics the input audio records. 

In this contribution, we explore how projections from latent acoustic representations can be decoded to optimize the parameters of a speech synthesizer. We compare three deep encoders and train simple neural nets to project their latent representations onto the parameters of the Pink Trombone (PT) model. The latent audio representations were obtained from trained and pre-trained models for speech. To restrict our analysis we focus on a restricted set of articulators and single- and multi-vowel records. 

The objectives for this paper include: \begin{itemize}
    \item \textit{Optimizing PT parameters} from latent representations' projections for encoder-decoder structures.
    \item \textit{Evaluating a two-head decoder VAE structure} on the PT parameters and audio reconstruction.
    \item \textit{Testing PT parameters' projection} from foundation encodings to validate our approach.
    \item \textit{Zero-shot testing of the latent representations} on human records while training on PT samples.
\end{itemize}

The remainder of this paper is organized as follows. Section 2 presents the Pink Trombone model and describes the statistical encoding approaches toward speech parameterization, including those derived from the pre-trained foundational audio encoder. Section 3 describes the dataset collected and the proposed coding of vocal articulations for optimizing the Pink Trombone parameters for speech synthesis. In sections 4 and 5, we present the results of our analysis and discuss them. Finally, in section 6, we summarize our conclusions.

\section{Related work}
\label{sec:sota}

Hereafter, we describe the PT synthesizer model and the different encoders evaluated. For each encoder model, we briefly discuss the decoding process to illustrate our investigation's relevance and challenges. 

\subsection{The Pink Trombone}

The PT model \cite{story2005parametric} provides a computational simulation for understanding speech production mechanics\footnote{\url{https://dood.al/pinktrombone/}}. Following the Kelly-Lochbaum technique, the model simulates the movement and interaction of key anatomical structures involved in speech production, such as the vocal cords, tongue, lips, and jaw. One can experiment with the variations in glottal excitation and vocal tract shape on the resonance characteristics of speech sounds. 

The model produces voice from a limited set of parameters that encode the vocal tract's configuration. These parameters include variations in the location and diameter of constrictions, energy and structure of the glottal excitation, etc. Users can adjust the model parameters to match the vocal tract resonances necessary for producing specific speech sounds. However, coordinating articulators to match a speech record is a complex task, as it is to identify the most suitable PT parameters. Contrarily, synthesizing audio records from the PT parameters is straightforward. 




\subsection{The Variational Autoencoder}
\label{subsec:VAE}

A Variational AutoEncoder (VAE) \cite{kingma2013auto} is a neural network used in unsupervised training. It comprises an encoder, which compresses input data, $x \in \mathbb{R}^E$, into a latent space representation, $z \in \mathbb{R}^D$, of lower dimensionality ($D \ll E$), and a decoder, which reconstructs the original input from the latent space. The latent space is learned through variational inference. The dimensionality of the latent space creates a bottleneck that makes data reconstruction a hard task, forcing the architecture to extract relevant features.

We use a $\beta$-VAE structure \cite{higgins2017beta} particularly suitable to leverage reconstruction and generative capabilities. The starting point is the assumption that the input data follows an actual probability distribution $p^*(x)$ that can be parametrically approximated: $p_{\vartheta}(x)$. By imposing a suitably simple prior distribution $p_{\vartheta,\varphi}(z)$ the parameters $\vartheta$ of the probabilistic decoder, $p_{\vartheta}(x|z)$, will induce a marginal distribution over the signal space.
To maximize the likelihood of the data, $\{x_i\}_{i=1,\ldots,N}$, we optimize the parameters $\vartheta$ to match the probabilistic distribution of a given set of observations (the dataset). Additionally, we aim to regularize the latent space to control the training process.

Computing $p_{\vartheta}(x)$ is analytically infeasible in most cases due to the intractability of $p_{\vartheta}(z|x)$. The solution to this problem is to avoid directly optimizing $p_{\vartheta}(x)$ and instead maximize a lower bound, known as the Evidence Lower Bound (ELBO). The training cost function is defined to train the parameters $(\vartheta,\varphi)$. It involves the two terms (Eq. \ref{eq:VLB}): one that represents the reconstruction error and the other that enhances the generative capabilities of the model. The control that the $\beta$-VAE provides over this trade-off is a key component of our methodology.
\begin{align}
\label{eq:VLB}
    \mathcal{L}_{\vartheta,\varphi}(x) = \!\!\!\!\!\!\!\!\underset{z \sim q_\varphi (z|x)}{\mathbb{E}} \!\!\!\!\!\! \!\!\left[
\log p(x|z) - \beta \!\cdot\! {\text D}_{\text KL} (
    q_{\varphi}(z|x) || p_{\vartheta}(z|x)) \right]
\end{align}
One may extend the previous cost function and incorporate additional terms to add new properties into the latent space. It ensures that the latent space displays natural clustering properties, mitigates the impact of local minima in training, or incorporates additional decoding capabilities. 

\subsubsection{Multi-head decoding}
\label{subsec:two-head_decoders}


The standard $\beta$-VAE model and cost function (Eq. \ref{eq:VLB}) can be extended, incorporating additional decoding heads and their corresponding costs. We propose a two-head decoder for the PT synthesizer input and output reconstruction. One decoding head focuses on reconstructing the audio record chunks. The other focuses on predicting the PT parameters that generated that chunk. This way, the VAE structure is generalized to address two goals: optimizing the synthesizer parameters and producing synthetic records that mimic vocal sounds. 

We aim to assess whether a dual-head approach enhances audio quality, the generative capabilities of the latent space, and computational speed for parameter generation. To prove this hypothesis, one of the heads might be removed for an ablation study. We can directly use a projector trained on the embeddings of an already pretrained model. The resulting audio quality will depend on the performance of the PT model. This method allows for training a projector on a previously established VAE or employing a different type of encoder (analysis head), which doesn’t have to be a VAE. While not guaranteeing that the latent representation maximizes input and output optimization, this setup simplifies the training process and benefits from existing high-performance encoder representations. 

\subsection{Pre-trained foundation models}

Neural network models trained on vast datasets learn general-purpose representations of input data, capturing valuable high-level features for various downstream tasks like classification, generation, and translation. Pre-trained foundational deep encoders play a crucial role in understanding and manipulating audio data in audio processing. Among these models, we aim to evaluate Wav2Vec \cite{NEURIPS2020_wav2vec} and EnCodec \cite{ACM_Encodec}. These are two encoders specifically designed to learn representations from raw audio.  

Wav2Vec employs self-supervised learning techniques to provide embeddings that capture meaningful features from speech signals and fine-tune them from transcribed speech. It enables practical tasks, e.g., speech recognition and speaker identification, in real-world conditions. When evaluated on speech recognition, Wav2vec 2.0 achieved a Word Error Rate of 1.8\% on the Librispeech dataset; significantly better than previous state-of-the-art models' 3.3\%. Wav2vec typically operates at a 16 kHz sampling rate, using 32 milliseconds window, and delivering a 768-dimension-long vector per frame.  

EnCodec focuses on compressing and reconstructing audio signals efficiently, a crucial aspect for applications like streaming and storage. Contrary to Wav2vec, EnCodec operates at 24 or 48 kHz sampling rate, uses multiple window lengths for speech analysis, and delivers 128-long encoded vectors.

\section{Materials and Methods}

\subsection{The PT dataset}
\label{sec:dataset}

We crafted three datasets comprising 10,000 1-second audio files each. Audio is sampled at 48kHz and synthesized using the PT. The synthesis process involves randomly sampling parameters within predefined ranges. We strictly focused in vocalic sounds, with particular attention to tongue-related articulation. The two parameters involved, constriction location and diameter, were interdependent and differently sampled to ensure adequate representation of vowels' acoustic space \cite{stevens1998acoustic}. The sampling of the tongue position followed a log-normal distribution, and its value restricted the range for the constrictions vocal tract area. All other parameters were restricted independently.

The three datasets were generated to capture various aspects of speech dynamics. The first focuses on static 1-second vowels, providing insight into steady-state phonetic characteristics. The second introduces linear parameter changes to simulate natural speech progression. The third incorporates faster variations by introducing random parameter sets every 100 milliseconds, facilitating exploration of rapid speech dynamics. Collectively, these versions offer a comprehensive exploration of speech characteristics and allow our models to behave well under different speech-like input signals.

After synthesis, the audio files were segmented into 15-millisecond windows, computed their 128-mel spectra and normalized. Windows were then shuffled for training while retaining information of the parameters that generated the current and the previous time windows. The goal was to preserve temporal continuity and contextual relevance (see Section \ref{subsec:two-head_decoding_VAE}). An 80-20 split between train and validation was also computed.

\subsection{The two-head decoding VAE}
\label{subsec:two-head_decoding_VAE}

As introduced in Section \ref{subsec:two-head_decoders}, the proposed two-head decoding VAE architecture is based on two main parts distinguished by their functionality. The first part is the encoder neural network, which processes input vocalic mel-spectrum vectors and produces lower dimensional latent representations which embed the acoustical information. These embeddings are fed to the decoding stage of the model, which is comprised of two heads (neural networks): the reconstruction and projection heads. While the task of the reconstruction head is to approximately invert the process undergone at the encoder, the projection head aims to predict the set of PT parameters ($\hat{\mathcal{P}}_{t}$) that "could have" produced the vocalic sound. Even though our interest is focused on the re-synthesis part (projection head), the reconstruction task  plays an essential role in conditioning the learning process. The resulting embeddings must be acoustically relevant, as well as physically informed.

On the one hand, the encoder and the reconstruction head are two one-dimensional Convolutional Neural Networks, with direct convolutions for the former and transposed convolutions for the latter. On the other hand, the projector is a feed-forward neural network which inputs the 64-dimensional embedding from the encoding stage and produces a 6-element PT parameter vector. This projector network is comprised of 4 layers. The ReLU activation function was used at the input and hidden layers, while output activations where chosen to be sigmoid functions to match normalization scheme of the data.

The proposed model requires the loss function to be multi-objective in its formulation (Eq. \ref{eq:loss_VAESynth}). \begin{equation}
\begin{split}
\label{eq:loss_VAESynth}
\mathcal{L}_{\vartheta,\varphi}(x,\mathbf{\hat{p}}_t) &= ELBO_{\vartheta,\varphi}(x) \!+\! \sum_{i=1}^{|\hat{\mathcal{P}}_{t}|} \beta_{t,i}\,(\hat{p}_{t,i} - p_{t,i})^2 \\
&+ \sum_{i=1}^{|\hat{\mathcal{P}}_{t}|} \beta_{t-1,i}\:\text{\textit{Huber}}(\hat{p}_{t,i},p_{t-1,i})
\end{split}
\end{equation}
The first term is the Evidence Lower Bound (ELBO), which corresponds to the mel reconstruction quality and latent space structure regularization (see Section \ref{subsec:VAE}, Eq. \ref{eq:VLB}). The ELBO depends on the parameters of the encoder ($\vartheta$) and decoder head ($\varphi$). The errors in the projection head are computed by estimating the current values for the PT parameters ($t$), taking into account previous ($t-1$) time windows. In the special case of the first window of the sound file, we simply repeated the parameter and acoustic information as if the vocal tract was static.

The loss for current window parameters is a squared euclidean distance, while the Huber loss \cite{Huber1964} penalizes abrupt parameter changes with respect to the previous time window. The rationale for this choice is that the considered behavior of the vocal tract will entail movements that can be captured using 15 ms frames \cite{stevens1998acoustic}. 
We also acknowledged that certain parameters ought to be more relevant, elastic, or conditioning than others in terms of the fidelity of the synthesis and smoothness. Thus, the loss for each current and prior parameter is weighted by a different $\beta$ factor (hyperparameters).

The proposed loss function favors the fact that latent embeddings contain relevant acoustical information from the synthesized speech (ELBO) while the estimation of the parameters of the synthesiser is accurate (euclidean distance) and smooth along time (Huber loss), all of which are desirable characteristics in a speech-to-articulation system. 



\subsection{Projection of foundation representations}

The neural network of the projector is designed to accept the 64-dimensional latent space of the VAE as its input. For EnCodec and Wav2vec, which have different dimensional embeddings, spherical interpolation has been used to adjust them to the desired dimension. This technique, well-documented in the literature, has been proven to preserve relevant acoustic information \cite{sanchez2024visual}.

\section{Results}

\subsection{Objective performance}

\begin{figure*}[t]
  \centering
  \includegraphics[width=\textwidth]{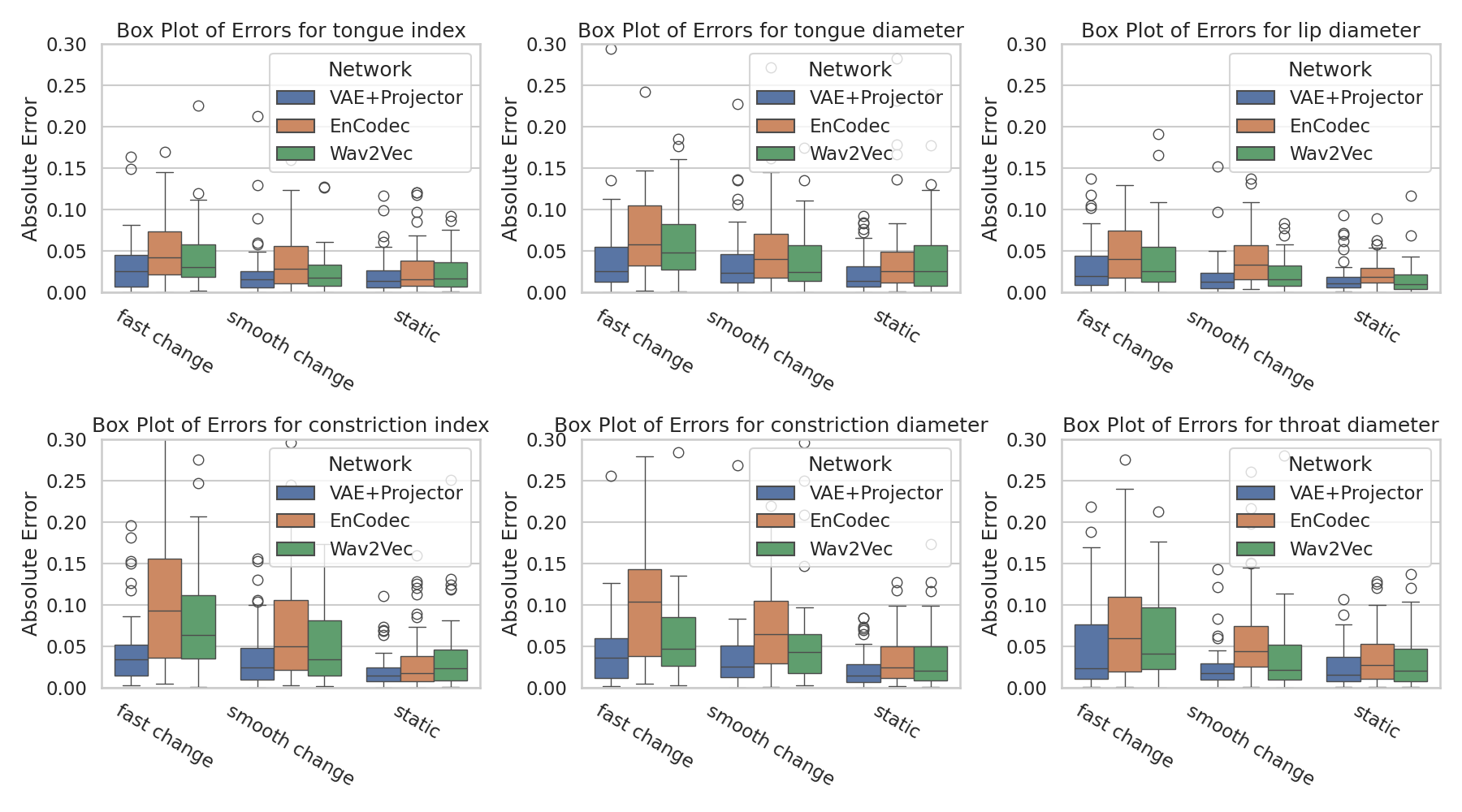}
  \caption{Absolute error of the predicted PT parameters for each of the neural models in each experiment: fadt change, smooth change, static. Errors are normalized between 0 and 1 for fair comparison.}
  \label{fig:objective}
\end{figure*}

Figure \ref{fig:objective} illustrates the normalized deviations in the estimation for each parameter in the validation subset. These represent the objective errors in network performance. We normalize the parameters to enhance comparability. Predicting parameters proved increasingly challenging as vocal tract movements accelerate. Static sounds exhibit lower errors compared to dynamic sounds, with rapid sounds incurring higher errors than their slower counterparts. This phenomenon occurs because variations in parameters expand the potential acoustic outcomes, intensifying with faster movements. Although this effect is minor, it is observable during network training.

Examining the parameters more closely reveals trends corresponding to their typology. For instance, parameters labeled "diameter" pertain to specific physical dimensions of the vocal tract that determine the extent of constriction (in cm). Conversely, "index" parameters primarily indicate the location of these constrictions. Intuitively, one may expect that the location is more elastic than the size. Shifting a constriction a few millimeters forward or backward in the vocal tract has less acoustic impact than altering the constriction’s opening, which changes the acoustic impedance significantly \cite{stevens1998acoustic}. This finding is consistent in the parameter data: parameters that significantly alter the acoustic signal are easier to predict as the optimizer in the neural network identifies clearer minima than in the indexes. Nonetheless, these objective errors do not necessarily correlate with subjective experiences, as larger parameter errors can still result in acoustically identical synthesized audios.

Comparing different models, the VAE+Projector, specifically trained for this task, surpasses foundation models in performance. This superiority stems from its customization for the specific task at hand. However, foundational models show remarkable robustness. Despite their original design as purely acoustic models, foundation models hold sufficient articulatory information within their latent spaces. On average, Wav2Vec records seem marginally lower errors than EnCodec, although the difference is not statistically significant. These disparities should relate to the distinct training regimens and datasets used for pre-training each model. From a computational standpoint, EnCodec operates faster than Wav2Vec as it is optimized for CPU execution, whereas Wav2Vec requires a GPU for optimal performance, presenting a trade-off between reconstruction quality and processing speed.

\subsection{Subjective-equivalent (ViSQOL) performance}

\begin{figure}[t]
\begin{center}
\includegraphics[width=\columnwidth]{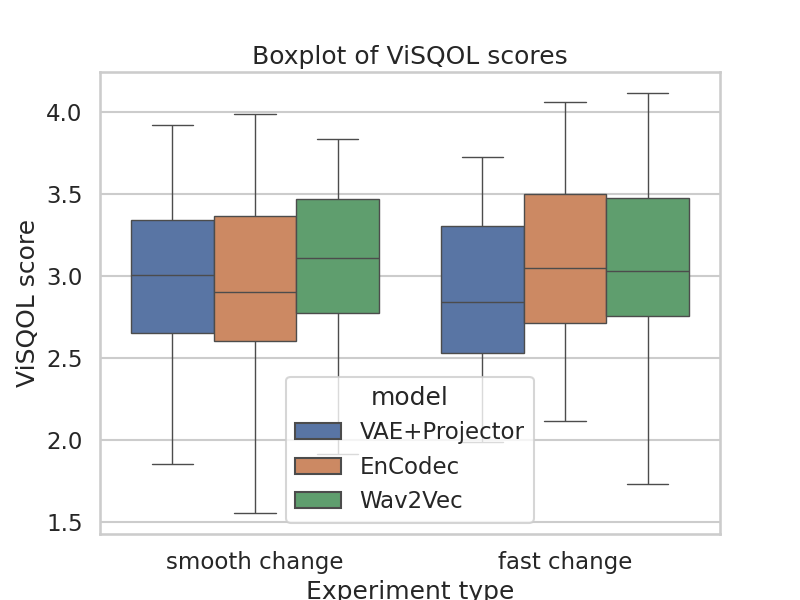}
\caption{ViSQOL scores for each of the neural models with respect to human audio recordings.}
\label{fg:visqol}
\end{center}
\end{figure}

We also evaluated audio quality computing the ViSQOL metric \cite{chinen2020visqol}. This metric uses the original audio reference to compare it with the generated record. The metric theoretically correlates with the subjective Mean Opinion Score (MOS) results. To enhance accuracy, we did not use the generic ViSQOL-to-MOS mapper; instead, we applied a mapper developed explicitly for the PT\footnote{\url{https://mateocamara.github.io/pink-trombone/#visqol-pink-trombone-svr-model}}.

Figure \ref{fg:visqol} displays the ViSQOL results for each model in dynamic experiments. We exclude static experiments because the analysis focuses on performance relative to human sounds. Though potentially static, humans inherently exhibit unavoidable variations due to articulators variations. Unlike the objective results previously presented, we do not observe clear superiority towards the trained model over foundational models in this case. This finding is particularly significant as it suggests that models never trained with PT sounds can still accurately mimic a person's original vocal tract positions. This supports the choice of PT parameters and even the decision to use the PT as a simple voice synthesizer. Moreover, these results indicate that the VAE+Projector training should not be overfitted to PT sounds, as it accepts human sounds and estimates articulatory parameters with equal fidelity.



\subsection{Ablation study}

\begin{figure}[t]
\begin{center}
\includegraphics[width=\columnwidth]{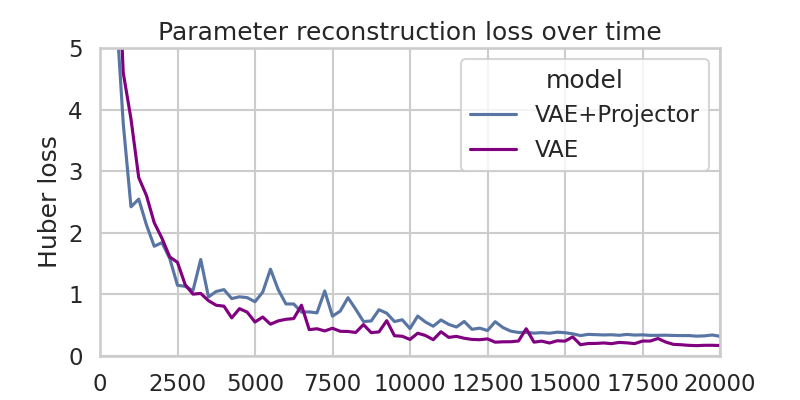}
\caption{Huber loss of parameter reconstruction during the training over the validation subset.}
\label{fg:param-rec}
\end{center}
\end{figure}

\begin{figure}[t]
\begin{center}
\includegraphics[width=\columnwidth]{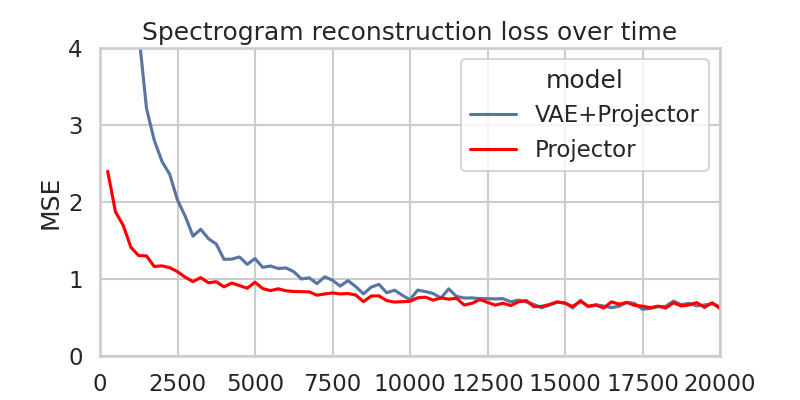}
\caption{MSE loss of mel reconstruction during training over the validation subset.}
\label{fg:spec-rec}
\end{center}
\end{figure}

To ensure that the VAE+Projector model was appropriately calibrated and did not tend towards overfitting on either of its two cost functions (spectrogram error and parameter error), we conducted an ablation study. In this experiment, we first analyzed the VAE network on its own. After training, we fixed its weights and then trained the projector. We compared this process with the integrated VAE+Projector network. Figures \ref{fg:param-rec} and \ref{fg:spec-rec} show the performance results on the validation subset during training.

The graphs indicate that although the validation error decreases more rapidly in the initial stages for the separately trained nets (projectors), the integrated training asymptotically converges to a similar point. This evolution is different for the Huber loss (Fig.~\ref{fg:param-rec}) and the MSE (Fig.~\ref{fg:spec-rec}). We conclude that the network weight optimizers have found a similar global minimum, proving that simultaneous training of the networks is effective. At the end of the simulation (20,000 epochs) the models seem to converge.

\subsection{Case of study for articulation: /ieaou/}

\begin{figure}[t]
\begin{center}
\includegraphics[width=\columnwidth]{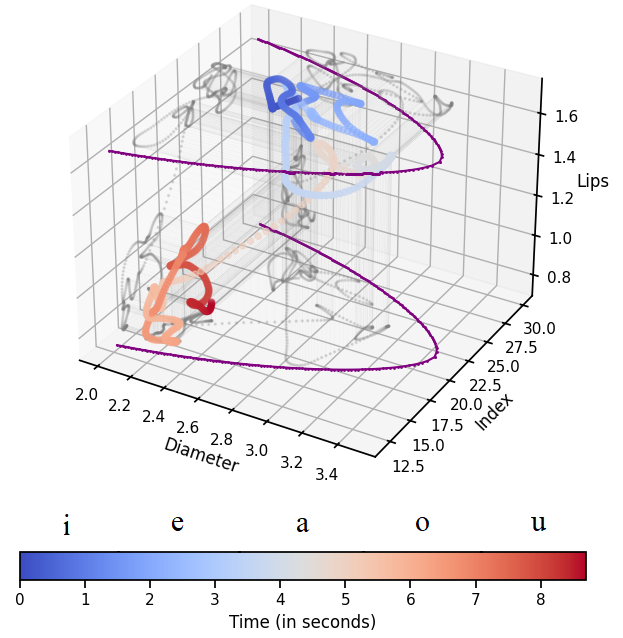}
\caption{3D view of the position of the tongue and lips for the /ieaou/ articulation. Purple lines are the limits of the tongue positions. The gradient of color (from blue to red) refers to the articulation over time. A more detailed plot can be found on the web resources attached.}
\label{fg:sincplot}
\end{center}
\end{figure}

The articulation of /ieaou/ in Spanish is a particular informative sequence. The transitions between articulatory positions predominantly involve changes in a single articulator at each moment, either the tongue or the yaw and the lips, with minimal involvement from other areas, such as the throat or the soft palate. Figure \ref{fg:sincplot} illustrates the evolution of the decoded parameters over time on a uttered record in a three-dimensional plot that depicts the values for two articulators and three PT parameters: two representing the tongue and one the lips.


The analysis can be broken down at each uttered vocal sound. Initially, /i/ is articulated with the tongue in a forward position at approximately (30, 2), indicating an advanced tongue position with the lips open. Following this, the /e/ sound keeps the mouth open while the tongue moves downward, represented as a movement from zone 2 to zone 3. Next, the /a/ requires the tongue to lift towards the rear of the palate, correctly depicted in the figure as moving to the area of (18, 2). Although the mouth should theoretically remain open, it starts closing, which is audibly evident in the recording. This is why the lips move from 1.6 towards the region of 1. The /o/ and /u/ sounds are quite similar; both show nearly closed lips, with the most notable differences observed in other PT parameters not displayed in this graph. These parameters indicate a tendency to close the entire vocal tract during /u/, producing a sound that lacks naturalness. 

We conclude that the synthesis of the first four sounds is realistic. Contrarily, have been the /u/ phoneme is not yet accurately replicated by our vocal tract model, though it is approaching accuracy. Future work will focus on improving this aspect.

\section{Discussion}

We highlight two significant aspects of our research. Firstly, it is fascinating to observe that large sound encoding models, although primarily trained for different tasks, contain articulatory information. This finding confirms that these models have effectively integrated sound within their embedding spaces, including physical details. This insight raises an interesting question for the field of neural voice synthesis: should well-known synthesizers be utilized to synthesize new audio based on neural encodings? Predicting parameters for a synthesizer offers advantages, such as ensuring a flawless reconstruction, which contrasts with potential issues from a neural decoder that often requires a vocoder and might introduce noise. 

Secondly, the results presented in this paper are comparable to those achieved with the PT synthesizer using pure optimization techniques. The PT is a non-differentiable optimizer that cannot be included directly in the autodifferentiable graphs of a neural network. As a result, an acoustic error function cannot be integrated directly into a neural network. However, our parametric cost function achieves results equivalent to those of acoustic cost functions, with one crucial difference: optimization is a lengthy and costly process, whereas neural network inference is remarkably fast. To illustrate this fact, optimizing one second of audio can take around 15 minutes, while neural network processing is nearly real-time. This efficiency is crucial for assessing the reconstruction quality achieved in this research, or to be integrated into a Digital Audio Workstation.

\section{Conclusions}

This research has successfully designed a neural system for encoding vowel sounds that predicts parameters for a voice synthesizer using a two-head decoding VAE. The complete system's performance, which compresses and projects audio into parameters, was compared with foundation encoders Wav2Vec and EnCodec. The results demonstrate a remarkable predictive capability for the considered articulatory parameters. When synthesizer sounds are fed into the network, our approach achieves a normalized median error of less than 0.1 in all cases. Additionally, the subjective-equivalent results with the ViSQOL metric, using human sounds as input, average a score of 3 on the MOS scale. This is significant value considering the regeneration occurs in a vocal synthesizer unrelated to the vocal tract of the actual speller. In terms of state-of-the-art contributions, this study allows rapid estimation of articulatory parameters on human vowel sounds, being three orders of magnitude faster than traditional optimizations \cite{camara2023optimization}.

Furthermore, multiple human sounds from different individuals with varying difficulty levels were analyzed. Numerous models were trained to assess reconstruction capabilities based on the original vocal tract's speed. Different sets of hyperparameters were also tested. This document presents those that led to the best reconstruction quality in the validation subset. Readers are invited to visit the website\footnote{\url{https://mateocamara.github.io/neural-pink-trombone/}} where they can find audio files, model weights, and other additional information.

Future work shall analyze new sets of articulatory parameters for any synthesizer to generate a wider variety of sounds, including more complex vowels and consonants. We also suggest introducing new cost functions to improve coherence with the signal being processed. Additionally, developing a differentiable version of the Pink Trombone to enable end-to-end acoustic cost functions would be beneficial, or alternatively, a differentiable model that predicts the potential acoustic error. 

\section{Acknowledgments}

Activities described in this contribution were partially funded by the European Union's Horizon 2020 Research and Innovation Programme under grant agreement No. 101003750, the Ministry of Economy and Competitiveness of Spain under grant PID2021-128469OB-I00.

\bibliographystyle{jaes}

\bibliography{refs}

\end{document}